\documentclass[12pt,preprint]{revtex4-1}
\usepackage{epsfig}

\usepackage{amsmath,amssymb,amsfonts,euscript}

\newcommand{\deriv}[2]{\,\mbox{$\displaystyle \dfrac{{\rm d}#1}{{\rm d}#2}$}\,}

\DeclareMathOperator{\arctg1}{arctg}

\begin{document}
	\renewcommand{\figurename}{FIG}
	\renewcommand{\tablename}{TABLE}
	
	\title{On the reconstruction of motion of a binary star moving in the external gravitational field of Kerr black hole by its redshift
	}

	\author{S. Komarov}
	\email[E-mail: ]{staskomarov@tut.by}
	\author{A. Gorbatsievich }
	\email[E-mail: ]{gorbatsievich@bsu.by}
	
	\affiliation{Theoretical Physics Department, Belarusian State University, Nezavisimosti av., 4, 220030 Minsk, Belarus}


\begin{abstract}
		We present a research of the time evolution of the redshift of light received from the binary star that moves in the external gravitational field of Kerr black hole. We formulate a method for the solution of inverse problem: calculating of the parameters of relative motion of stars in binary system using the redshift data. The considered formalism has no restrictions on the character of the motion of the center of mass of a compact binary star and can be applied even in the case of binary motion close to the event horizon of a supermassive black hole. The efficiency of the method is illustrated on a numerical model with plausible parameters for the binary systems and for the supermassive black hole, which is located in the center of our Galaxy. 
\end{abstract}
\nopagebreak

\maketitle

\section{Introduction}

In recent years the problem of calculating of radiation characteristics of a source that moves in external gravitation field of Schwarzschild or Kerr black hole is considered in many papers (see, e. g., \cite{Zhang2015, GrouldS2, Tarasenko,  Psaltis2016, Zhang2017, Angelil}). On the one hand it provides possibilities for studying our Universe. On the other hand these researches can be useful for future tests of different theories of gravity. 

One of the interesting results that are provided by such studies is the possibility of solution of inverse problem: unique determination of parameters of the source, its motion and external gravitational field using the data from received electromagnetic radiation. Moreover, underlying mathematical models that are used in the works \cite{Zhang2015, GrouldS2, Tarasenko,  Psaltis2016, Zhang2017, Angelil} enable to investigate the effects of strong gravitational field on the motion of the source and propagation of its radiation. The inverse problem for the source moving in the equatorial plane of a Kerr black hole was solved in \cite{KerrRedShift} using the redshift of the source only. The case of more general motion of the source in gravitational field of non rotating supermassive black hole is considered in \cite{Tarasenko}, but in the cited paper the redshift data as well as the radiation intensity both are used for the solution of inverse problem.

In the present paper we study the redshift of light  that is emitted by a binary star moving in the external gravitational field of Kerr black hole. Unlike other radiation characteristics, the redshift of spectral lines of a star can be measured with a very hight accuracy (see, e. g., \cite{Gillessen2010}) that gives possibility to study general relativistic effects in the motion of certain  currently observed stars (see, e. g., \cite{GrouldS2}). In the previous paper of the authors \cite{GRG2018} it was shown that the problem of calculating of the redshift of light received from compact binary star system mathematically can be considered as two independent problems. They are correspond to the motion of the center of mass of the system and to the relative motion of its components. Due to the first problem is widely studied in literature (see, e. g., \cite{Zhang2015, KerrRedShift}), we focus only on the case of the determination of the parameters of relative motion when the 
motion of the center of mass and the parameters of external Kerr gravitational field are known.

It is necessary to take into account the influence of strong gravitational field not only on the motion of the source but also on the worldline of emitted light.  All these effects can be important for testing theories of gravity. Due to this we safe full general relativistic treatment for the consideration of light propagation in space-time (in geometrical optics approximation) and for the motion of the stars. The considered method have no restrictions on the motion of the binary as a whole. The trajectory of the motion can be even close to event horizon of a supermassive black hole and velocity of the center of mass of the binary can be relativistic or ultrarelativistic. We will consider the redshift in a linear approximation relative to $v$ and $\rho$, where $v$ is the velocity of the components of the binary star relative to each other (in the co-moving reference frame) and $\rho$ is the distance between the two stars. We will show that with the additional assumption when the mass of the emitted source is known, it is possible to calculate all the parameters of relative motion of the stars with certain accuracy.

The problem that is considered in the present paper can be interesting for the purpose of studying of compact binary stars moving in the vicinity of the Galactic Center supermassive black hole (see, e. g., \cite{Gillessen2017, GBHL, Gillessen2010, NatureXsources, GrouldS2}).         

\section{Mathematical model of the source in binary star that moves in external gravitational field of Kerr black hole}\label{Model}
Consider a binary star that freely moves in external gravitational field of a black hole. Let one of the components of the binary be a source of electromagnetic radiation, that can be received by the observer on Earth. The system satisfies the following conditions (see earlier work of the authors for more detailed description of the system \cite{GRG2018}):
\begin{enumerate}
\item Mean distance between the stars is much greater than their own sizes which means that we can consider them
being mass-points.  Thus,
 we neglect also the effect of the proper rotation of the stars on their motion.\label{As1}
\item Motion of the stars relative to each other is non-relativistic, because the energy of their gravitational interaction with each other is much less than the rest energy.\label{As2}
\item The scale of inhomogeneity of external field is greater than the size of the system.\label{As3}
\end{enumerate}
In \cite{GRG2018} was shown that in the framework of formulated conditions, the motion of the center of mass of the binary satisfies Mathisson-Papapetrou equations. As a rule, for real binary star interaction of its proper angular momentum with the curvature of space-time is very small and motion of the binary can be considered as geodesic with good accuracy.   

In order to describe the relative motion of the stars it is convenient to use the co-moving Fermi coordinates of the observer  who can be placed in the center of mass of the system \cite{Mitskievich,Mi,Fortini}. Consider world line $x^i=\xi^i(\tau)$ of this observer, where $\tau$ is the proper time of his.  Along $\xi^i(\tau)$ we define co-moving tetrad (vierbein) $h_{(m)}{}^i$:
\begin{equation}
h_{(4)}{}^i=\frac{1}{c}u^i, \quad h_{(i)}{}^k h_{(j)k}=\eta_{(i)(j)}\,,
\end{equation}
where $u^i=\mathrm{d}\xi^i/\mathrm{d}\tau$ is 4-velocity of the observer, $\eta_{(m)(n)}=\mathrm{diag}(1,1,1,-1)$, $c$ is the speed of light.
\begin{figure}
	\includegraphics[width=0.7\textwidth]{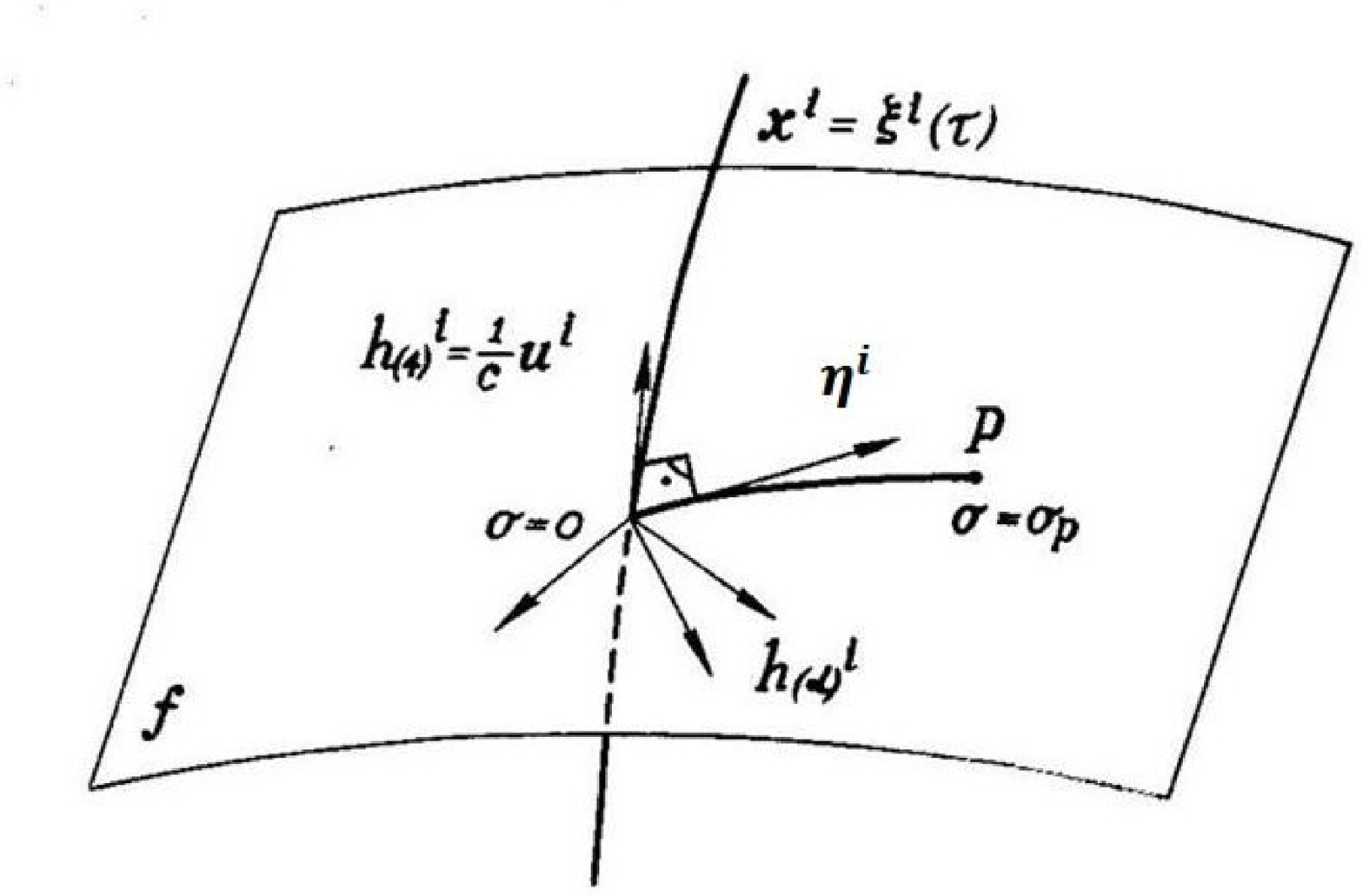}\\
	\caption{To the definition of generalized  Fermi coordinates}\label{fermi}
\end{figure}
Any point in the vicinity of worldline $\xi^i(\tau)$ can be given coordinates $\{x^{\hat i}\}$\footnote{Latin indices run from 1 to 4; Greek indices run from 1 to 3; signature of space-time metric is +2.}  in the following way (see Fig.~\ref{fermi}).
At first, we construct a spacelike geodesic hypersurface, that is orthogonal to worldline $\xi^i(\tau)$ at the point $O$ which belongs to
$\xi^i(\tau)$.
All points on this hypersurface have a coordinate $x^{\hat{4}}=c\tau$.
Then we find a geodesic line, which goes through $O$ and any point $P$ lying in the hypersurface.
At the point $O$ we construct a tangent vector to the geodesic.
Finally, we assign to P three coordinates $X^{(\alpha)}=\sigma_Ph^{(\alpha)}{\!}_i\,\eta^i$,
where $\sigma_P$ is the canonical parameter of the geodesic, evaluated at the point $P$. 

The equation of the relative motion of the components of the binary in co-moving Fermi coordinates for with $\xi^i(\tau)$ coincides with the worldline of the center of mass of the binary has the following form \cite{GRG2018}:
\begin{equation}
\begin{split}
&\frac{dv^{(\kappa)}}{d\tau}=\left(\frac{G(m_{1}+m_{2})}{r}\right)_{,(\kappa)}-2\varepsilon^{(\kappa)}{}_{(\alpha)(\tau)}\omega^{(\alpha)}v^{(\tau)}-\\
&\frac{2c(m_{2}-m_{1})}{(m_{1}+ m_{2})}R^{(\kappa)}{}_{(\nu)(\mu)(4)}x^{(\mu)}v^{(\nu)}+2D^{(\kappa)}{}_{(\nu)}x^{(\nu)}\,.\label{A29}
\end{split}
\end{equation}
Here $m_{1,2}$  are the masses of the components of the binary, $X^{(\alpha)}_{1,2}$ are Fermi coordinates of the components of the binary with respect to its center of mass and $v^{{\alpha}}_{1,2}$ are their velocities, $x^{(\alpha)}=X_2^{(\alpha)}-X_1^{(\alpha)}$ are Fermi coordinates of the relative position of the components. $\omega^{(\alpha)}$ is the angular velocity of the tetrad:
\begin{equation}\notag
\omega^{(\alpha)}=\frac{1}{2}\varepsilon^{(\alpha)(\kappa)(\tau)}h_{(\tau) i}\frac{Dh_{(\kappa)}{}^{i}}{D\tau}.
\end{equation}
Also the following abbreviation is introduced:
$$D_{(\mu)(\nu)}=-\frac{c^2}{2}R_{(4)(\mu)(4)(\nu)}+\frac{1}{2}(\delta_{(\mu)(\nu)}\omega^{2}-\omega_{(\mu)}\omega_{(\nu)}).$$
$\varepsilon^{(\alpha)}{}_{(\beta)(\gamma)}$ is the Livi-Chevita symbol, tetrad components of the curvature tensor $R_{ijkl}$ can be calculated as
$R_{(m)(n)(p)(q)}=h^i{}_{(m)}h^j{}_{(n)}h^k{}_{(p)}h^l{}_{(q)}R_{ijkl}$.

In the geometrical optics approximation (see, e. g., \cite{Stephani}) the radiation of the star propagates along an isotropic geodesic. The tangent vector $k^i$ to this geodesic is the wave vector of the light and it satisfies the following relations $k^ik_i=0,$ $k_{i;j}k^j=0.$ Consider the external gravitational field with time like Killing vector field $\mu^i$ (on spatial infinity $\mu^j\mu_j=-1$). The redshift $z$ of light that is received by stationary observer ($(u^i)_o=c\mu^i$, where $(u^i)_o$ is the velocity of the observer) can be found from (see, e. g., \cite{Zhang2017}) 
\begin{align}\label{generalredshift1}
z=\delta\lambda/\lambda=\frac{1}{\omega_0}(k^i)_s(u_i)_s-1\,.
\end{align} 
Here $\lambda$ is the wavelength of the emitted light, $\delta\lambda$ is the difference between the wavelengths of the received and the emitted light. $(u_i)_s$ is the 4-velocity of the source, $(k^i)_s$ is the wave vector of  light in the point of emission and $\omega_0=ck^i\mu_i$ (see, e. g., \cite{GRG2018}). For the source in a binary system the following relation holds (see \cite{GRG2018}):
\begin{align}\label{zz1}
z(\tau)=(1+z_0(\tau))\left(1-\frac{1}{c}\deriv{}{\tau}(n_{(\alpha)}X_1^{(\alpha)})\right)-1+\mathrm{O}\left(\frac{\varrho^2}{M^2}, \frac{v^2}{c^2}\right)\,,
\end{align} 
where $z_0(\tau)$ is the redshift of light of the imagine source that is located in the center of mass of the binary, $\tau$ is the proper time of the center of mass of the binary, $n^{(\alpha)}=k^{(\alpha)}/\sqrt{k_{(\beta)}k^{(\beta)}}$ is the normalised 3-wave vector of the light ray in the point of radiation. All the functions of time in (\ref{zz1}) must be calculated on the space-like hypersurfaces $\tau=const$ that are orthogonal to geodesic $(u^i)_s(\tau)$ for all $\tau$. In order to obtain the redshift as a function of time of observation, it is necessary to express the proper time (\ref{zz1}) in terms of time of observation $t$:
\begin{equation}\label{time}
t(\tau)=\int\limits^{\tau}_0(1+z(\tau'))\mathrm{d}\tau'\,.
\end{equation}
Inversion of formula (\ref{time}) gives the function $z(\tau(t))=z(t).$

\begin{figure}[h!]
\includegraphics[width=\linewidth]{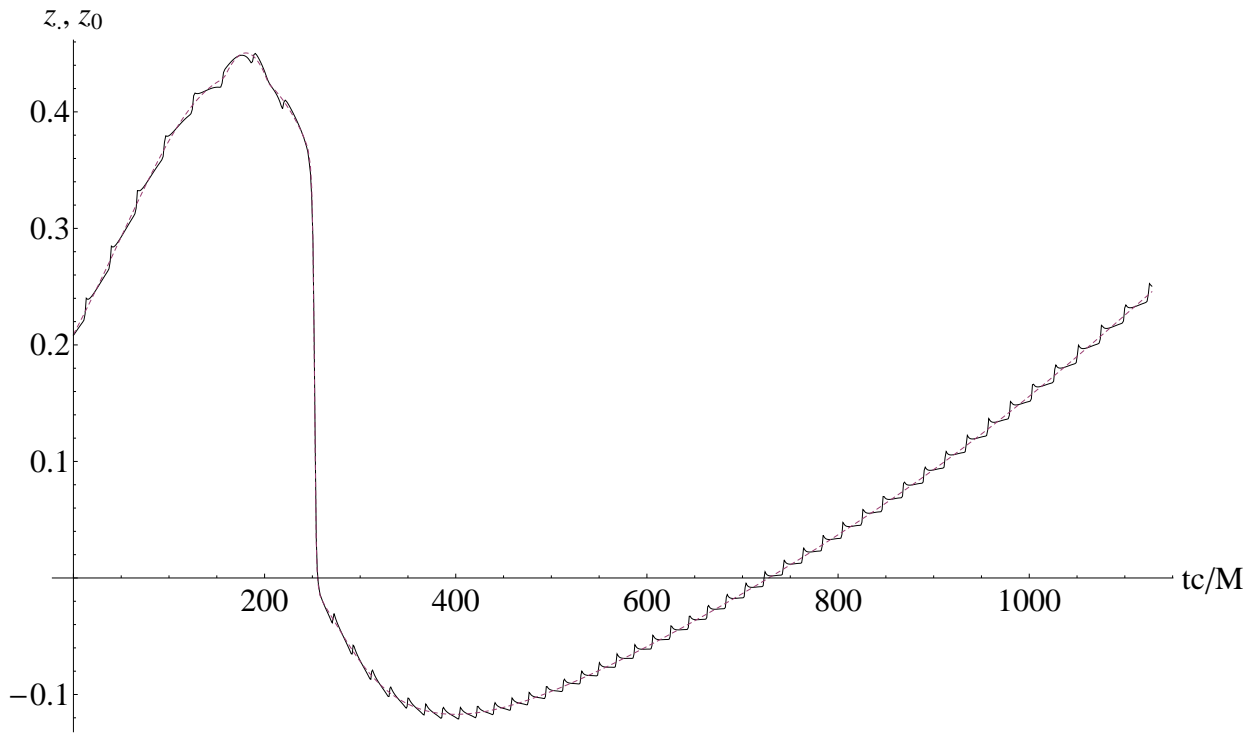}
\caption{Redshift $z(t)$ (solid) and $z_0(t)$ (dashed) as functions of time of observation $t$. Parameters of motion are following: Kerr parameter $a=0,1 M;$ angular momentum per unit mass, $L=4,9 M;$ Carter constant per unit mass $Q=3,0 M;$ the energy per unit mass, $E=0,984;$  the mass of the source, $m_1=8,89\cdot 10^{-7}M;$ the mass of the companion star, $m_2=4,45\cdot 10^{-7}M;$ the binitial relative position, $x^{(\alpha)}(0)=\{0; 0,04 M; 0\};$ the initial relative velocity, $v^{(\alpha)}(0)/Mc=\{0,002; 0; 0,0015\}.$}\label{tKerrRedShift_a01}
\end{figure}

\begin{figure}[h!]
	\includegraphics[width=0.7\linewidth]{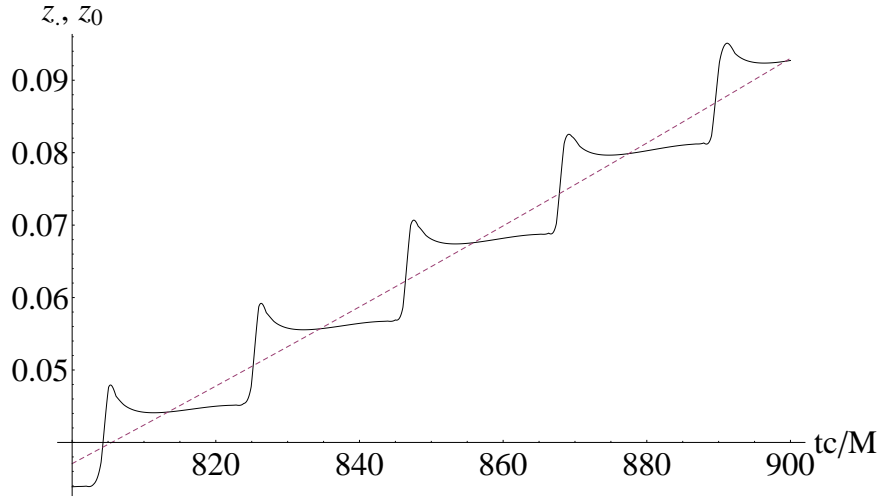}
	\caption{Magnificated part of the graphics of $z(t),$ $z_0(t)$ on Fig. \ref{tKerrRedShift_a01}}\label{tsmallKerrRedShift_a01}
\end{figure}

The supermassive black hole in the center of our Galaxy must not have physically relevant value of the electric charge. Due to this the most general metric that describes gravitational field of the supermassive black hole is the Kerr metric. In the present work we focus on this case of external gravitational field.
An example of redshift for a binary star in external Kerr field that has been calculated for certain model parameters satisfying assumptions about the system \ref{As1}-\ref{As3} is presented on Fig. \ref{tKerrRedShift_a01}, \ref{tsmallKerrRedShift_a01}. In the calculation of analytical expressions for isotropic geodesics in this example the linear approximation in Kerr parameter $a/M$ is used (see \cite{DAN2018}), where $a$ is the angular momentum of the black hole, $M$ its mass in geometrical units. We choose $M=4\cdot 10^6 M_{\odot}$, where $M_{\odot}$ --- mass of the Sun. Order of this value corresponds to the data for mass of supermassive black hole in the Galactic Center (see, e. g., \cite{Gillessen2017, GBHL, Gillessen2010}).
 
\section{Inverse problem}
The solutions of the problem of determination of the redshift as a function of time of observation for extended object in external gravitational field was described in the previous section. In the present section we consider a problem of determination of the parameters of binary system that moves in external gravitational field using the redshift of light, that is emitted by a star in binary, as a function of time of observation. We will refer to the last problem as the inverse problem.

\subsection{Decomposition of redshift}
\label{SecDec}
The observed redshift $z(t)$ depends on the large number of parameters which determine the trajectory of motion of the center of mass of the system and the trajectory of the relative motion of  the components and its masses. Due to this it seems not possible to obtain simple formulas for the unique exact solution of the inverse problem. One can expect obtain only approximate expressions. We propose rather more simple approach that gives possibilities to obtain parameters of motion of binary star with fixed accuracy. Moreover the present method can be generalized in order to obtain more high accuracy. In order to simplify the inverse problem it is possible to split the whole redshift $z(t)$ into two parts: the redshift for the center of mass $z_0(t)$ and the redshift for the relative motion of the components of the system $z_r(t)$. This can be performed in the following way. 

Assume that from the observation only the function $z(t)$ is known. It consists of two parts: slowly changing $z_0(t)$ and quickly oscillating part $z_r(t)=\mathrm{d}(n_{(\alpha)}X_1^{(\alpha)})/(c\mathrm{d}\tau).$ Consider an empiric function $f(t)\approx z_0(t)+1.$  From (\ref{zz1}) follows:
\begin{equation}\label{intz}
c\int\limits_{T_1}^{T_2}\left(\frac{1+z(\tau_P)}{f(t(\tau_P))}-1\right)\mathrm{d}\tau_P\approx\left.n_{(\alpha)}X_1^{(\alpha)}\right|^{T_1}_{T_2}+c(T_2-T_1)\cdot\mathrm{O}\left(\frac{\varrho^2}{M^2}, \frac{v^2}{c^2}\right)\,.
\end{equation}
In the approximation in which terms $\mathrm{O}\left(\varrho^2/M^2, v^2/c^2\right)$ can be neglected, equations of relative motion (\ref{A29}) formally coincide with the Newtonian equations for the binary in rotating coordinate frame. The accuracy of the solution for this approximation has the order of $c\tau\cdot \mathrm{O}\left(\varrho^2/M^2, v^2/c^2\right)$ (that follows from the equations of relative motion (\ref{A29})), that in fact coincides with accuracy of the relation (\ref{intz}). Due to this we will use this approximation for description trajectory of the relative motion of stars in Fermi coordinates.

\begin{figure}[h!]
\includegraphics[width=0.8\linewidth]{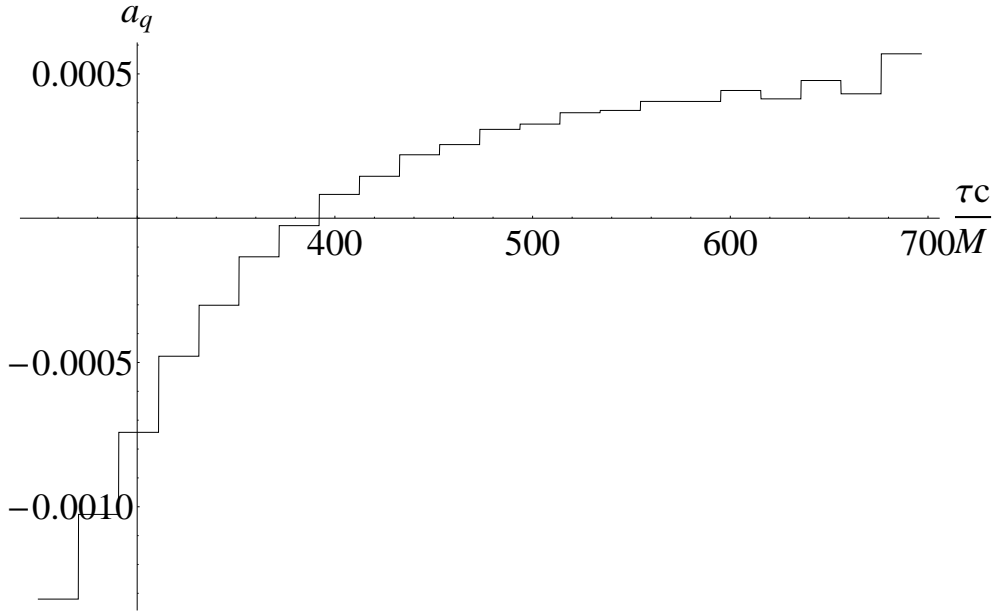}
\caption{Coeffitients $a_{q(\tau)}$ as functions of proper time of the center of mass of the binary. The function $q(\tau)$ is defined as the whole part of the proper time $\tau.$ The corresponding values $a_q$ obtained as solution of (\ref{eq2}) with the parameter $a_0=-0,00132$. This close to (\ref{a0}).}\label{aqKerr}
\end{figure}

\begin{figure}[h!]
\includegraphics[width=\linewidth]{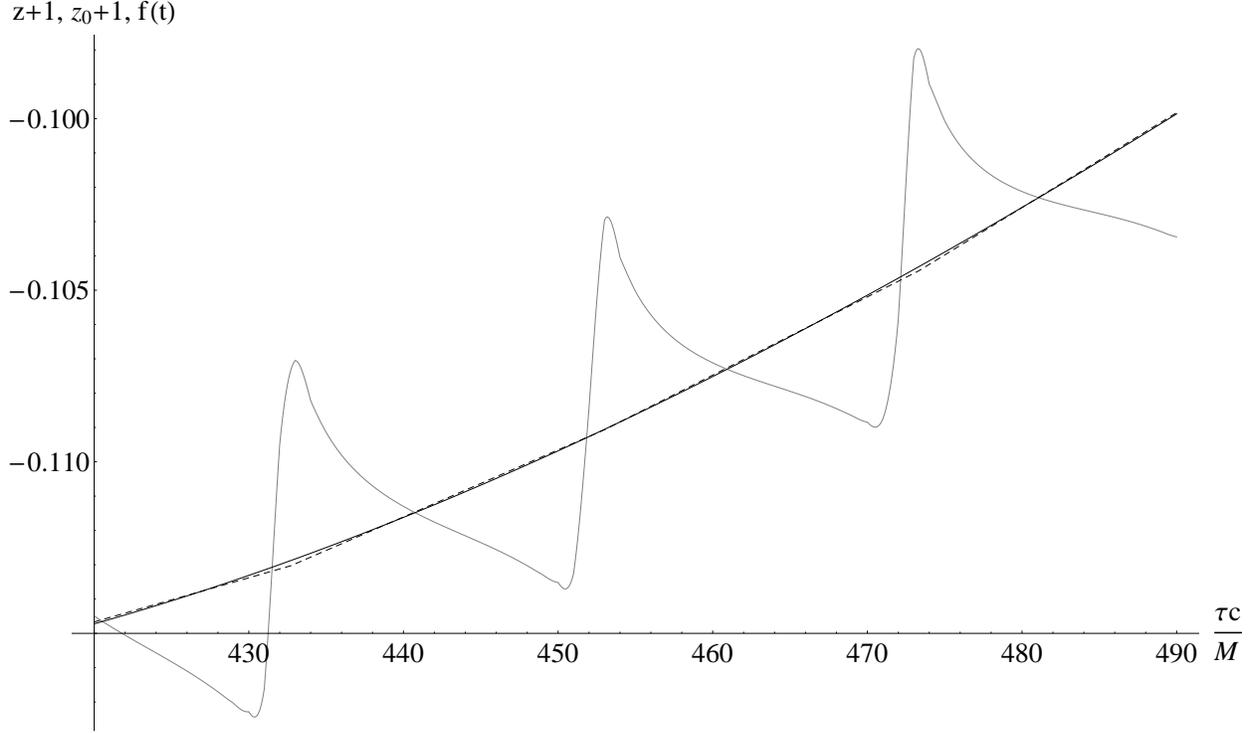}
\caption{Functions $f(\tau)$ (dashed), $z_0(\tau)$ (solid) and $z(\tau)$ (grey), where $\tau$ is the proper time of the center of mass of the system.}\label{DecRelativeMotionKerr}
\end{figure}

Furthermore the expression for the part of redshift that is determined by relative motion $z_r$ is formally coincide with the redshift in Newtonian mechanics. It is known from the expression for Newtonian redshift (see, e. g., \cite{Cherepaschuc}) that the points for with $n^{(\alpha)}X_{1(\alpha)}=0$ are extremums of function $z_r(\tau).$ This is true in our case if we neglect the derivative of unit 3-wave vector of emitted light with respect to non rotating Fermi frame:  
\begin{equation}\label{B}
B_{(\alpha)}=e_{(\gamma)(\beta)(\alpha)}n^{(\gamma)}\omega^{(\beta)}+\dot{n}_{(\alpha)}\,,
\end{equation}
where $B_{(\alpha)}\sim 1/T_0$ is small quantity. Here we introduce $T_0$ as characteristic time of changing of function $z_0.$ For certainty it can be calculated as proper period of radial coordinate of motion of the center of mass $r(\tau).$ Then we can chose times $T_2$ and $T_1$ such that they are coincide with certain extremum of $z(t)$ (minimum for certainty) and $T_2-T_1=nT$.  Here $T$ is the period of corresponding Newtonian motion and $n$ is a natural number. With sufficient accuracy $T$ can be found from the function $z(\tau)$. Then
\begin{equation}\label{nx}
\left.n_{(\alpha)}X_1^{(\alpha)}\right|^{T_1}_{T_1+nT}=ncT\cdot\mathrm{O}\left(\frac{\varrho^2}{M^2}, \frac{v^2}{c^2}\right)\,.
\end{equation}
Chose $n=1,$ $t_q=t(T_1+qT)$ and from (\ref{intz}), (\ref{nx}) obtain the system of $N+1$ equations:
\begin{equation}\label{eqf}
\int\limits^{t_{q+1}}_{t_q}\frac{\mathrm{d}t}{f(t)}\approx T, \text{ for }q=0,\,1,...N\,.
\end{equation}
In order to determine the function $f(t)$ in the unique way it is necessary to define additional properties of this function. Chose it as continuous function that is linear on the each interval $(t_q,\,t_{q+1})$. From (\ref{eqf}) obtain
\begin{align}
f(t)=a_q t+b_q, \quad t\in[t_q,\,t_{q+1}]\,;\notag\\
\left\{
\begin{aligned}\label{systemz_0}
&a_qt_{q+1}+b_q=a_{q+1}t_{q+1}+b_{q+1}\,;\\
&\int\limits_{t_q}^{t_{q+1}}\frac{\mathrm{d}t}{a_q t+b_q}=T\,.
\end{aligned}
\right.
\end{align}  
The solution of (\ref{systemz_0}) has the form
\begin{align}\label{eq1}
&f(t)=a_q t+a_q\frac{t_{q+1}-t_q\exp{(a_q T)}}{\exp{(a_q T)}-1}\,,
\end{align}
where $a_q$ can be found as a solution of the following recursive equation:
\begin{align}\label{eq2}
&a_q (t_{q+1}-t_q)\frac{\exp{(a_q T)}}{\exp{(a_q T)}-1}=\frac{a_{q+1} (t_{q+2}-t_{q+1})}{\exp{(a_{q+1}T)}-1}\,.
\end{align} 
It follows from the equations (\ref{eq1}) and (\ref{eq2}) that the function $f(t)$ has one undetermined parameter $a_0.$ It can be found from the additional assumption. It is known that function $z_0(t)$ slowly changes with time. Due to this we will find such parameter $a_0$ that function $f(t)$ is maximally smooth in some sense. We minimised the sum $\sum\limits_{q=0}^{N-1}(a_{q+1}-a_q)^2$ numerically. In order to find the initial value for the estimation of $a_0,$ consider an approximate solution of (\ref{eq2})
\begin{equation}\label{approxaq}
a_{q+1}=\frac{\sqrt{3}}{2T}\left(\sqrt{8\frac{t_{q+2}-t_{q+1}}{t_{q+1}-t_q}-5}-\sqrt{3}\right)-a_q+\mathrm{O}\left((a_qT)^2\right)\,.
\end{equation} 
$a_q$ has the sense of time derivative of $z_0(t)$ and has the order of $1/T_0$. Then we have $a_q T\sim T/T_0$ --- is a small parameter. The concrete value of $T/T_0$ can be found directly from the function $z(\tau).$ This is convenient for the determination of the exact value of the approximation. It is follows from (\ref{approxaq}) that the graphics of functions $a_q(a_0)$ approximately represent two sets of straight lines with inclination angles $\pi/4$ and $-\pi/4$ respectively. Graphic of $a_0(a_0)$ intersect graphic of $a_1(a_0)$ near the value of $a_0$:
\begin{equation}\label{a0}
a_0=\frac{\sqrt{3}}{4T}\left(\sqrt{8\frac{t_{2}-t_{1}}{t_{1}-t_0}-5}-\sqrt{3}\right)\,.
\end{equation}     
Due to this the true value of $a_0$ must approximately coincide with (\ref{a0}). More exact value of $a_0$ can be obtained by analysing sequence of $a_q.$ Numerical results for calculating $a_q$ are presented in Fig. \ref{aqKerr}. We calculate parameters of the function $f(t)$ choosing $T_1=400$ and $N=22$. The graphic of obtained function is presented on Fig. \ref{DecRelativeMotionKerr}.  
\subsection{Inverse problem for the relative motion of the components}
\label{SecRel}
After the application of methods that are described in Section \ref{SecDec} two functions are known:
\begin{equation}
z_0(t)\approx f(t)\text{ and }z_r(t)\approx 1-\frac{z(t)+1}{f(t)+1}\,.
\end{equation}
Function $z_0(t)$ is determined only by the parameters of motion of the center of mass (in the approximation when $ct\cdot \mathrm{O}(v^2/c^2,\rho^2/M^2)$ can be neglected). The problem of the extracting of mentioned parameters from the function $z_0(t)$ is analogous to the problem of the extracting parameters of motion of a single source in external gravitational field. Some authors have proposed methods for the solution of such problem (see, e. g., \cite{KerrRedShift}, \cite{Zhang2015}). In the present work we consider only the inverse problem for the relative motion that is determined by function $z_r(t).$ We assume that the parameters of motion of the center of mass of the system are known. Because of this $n^{(\alpha)}$ and $\omega^{(\alpha)}$ are definite functions of proper time.

Consider the case when radiation from only one component (with mass $m_1$) of the binary system can be received by the observer on Earth. In order to use formulas (\ref{A29}) and (\ref{zz1}) for the motion of the stars and the redshift it necessary to introduce a Fermi basis that satisfy standard relations \cite{Mitskievich,Mi,Fortini}.  Numerical calculations in the present work are performed for the following Fermi tetrad in external Kerr gravitational field (due to the complexity of the expressions (\ref{h1}), (\ref{h2}), here we use the system of units for with $M=c=1$):
\begin{align}\label{h1}
&h^i{}_{(1)}=\left\{0,\,0\,,1\,,L/E\right\}(1/\sqrt{\EuScript{N}_1})\,;\\\notag
&h^i{}_{(2)}=\left\{1,\,0\,,\EuScript{M}_3\,,\EuScript{M}_4\right\}(1/\sqrt{\EuScript{N}_2})\,;\\\notag
&h^i{}_{(3)}=\left\{1,\,\EuScript{F}\,,\EuScript{M}_1\,,\EuScript{M}_2\right\}(1/\sqrt{\EuScript{N}_3})\,;\\\notag
&h^i{}_{(4)}=\left\{\EuScript{U}_1,\,\EuScript{U}_2\,,\EuScript{U}_3\,,\EuScript{U}_4\right\}\,,
\end{align}
where the following abbreviations are introduced:
\begin{align}\label{h2}
&\EuScript{P}=a^2 \left(E^2-1\right) \cos (4\theta)+a^2\left(1-E^2\right)+4\cos(2\theta)\left(L^2+Q\right)+4L^2-4Q\,;\\\notag
&\EuScript{M}_1= \frac{2arE + ((\rho^2 - 2r)L)/\sin^2\theta}{\Delta\sqrt{((r^2 + a^2)E - aL)^2 - \Delta (Q + (L - aE)^2 + r^2)}}\,;\\\notag
&\EuScript{M}_2=\frac{E\Sigma^2-2aLr}{\Delta\sqrt{\left(E\left(a^2+r^2\right)-aL\right)^2-\Delta \left((L-aE)^2+Q+r^2\right)}}\,;\\\notag
&\EuScript{M}_3=\left(2aEr\sin ^2\theta+L(\rho^2-2r)\right)\times\\\notag
&\frac{\sqrt{\left(E \left(a^2+r^2\right)-aL\right)^2-\Delta\left((L-aE)^2+Q+r^2\right)}}{\Delta\left(-4aELr\sin^2\theta+E^2\Sigma^2 \sin^2\theta-L^2(\rho^2-2 r)\right)}\,;\\\notag
&\EuScript{M}_4=\frac{\sin^2\theta(E\Sigma^2-2aLr)\sqrt{\left(E\left(a^2+r^2\right)-aL\right)^2-\Delta\left((L-aE)^2+Q+r^2\right)}}{\Delta\left(-4aELr\sin^2\theta+E^2\Sigma^2\sin^2\theta-L^2(\rho^2-2r)\right)}\,;\\\notag
&\EuScript{U}_1=\frac{1}{\rho^2}\sqrt{\left(E\left(a^2+r^2\right)-aL\right)^2-\Delta\left((L-aE)^2+Q+r^2\right)}\,;\\\notag
&\EuScript{U}_2=\frac{1}{\rho^2}\sqrt{Q-\cos^2\theta\left(a^2 \left(1-E^2\right)+L^2/\sin^2\theta\right)}\,;\\\notag
&\EuScript{U}_3=(2aEr+L(\rho^2-2r)/\sin^2\theta)/(\Delta\rho^2)\,;\\\notag
&\EuScript{U}_4=(E\Sigma^2-2aLr)/(\Delta\rho^2)\,;\\\notag
&\EuScript{F}=\frac{-\sqrt{\left(E\left(a^2+r^2\right)-aL\right)^2-\Delta\left((L-aE)^2+Q+r^2\right)}+\Delta\left(E\EuScript{M}_2-L\EuScript{M}_1\right)}{\Delta  \sqrt{Q-\cos^2\theta\left(a^2\left(1-E^2\right)+L^2/\sin ^2\theta\right)}}\,.
\end{align} 
Here $\{r,\,\theta,\,\phi,\,t\}$ are Bouer-Lindquist coordinates, $E=(u_i)_s\mu^i$ is the integral of motion along timelike geodesic of the center of mass (with tangent vector $(u_i)_s$) of the system that is associated with the timelike Killing vector field $\mu^i.$ And integral of motion $L=(u_i)_s\psi^i$ associated with the Killing vector $\psi^i=(\partial/\partial{\phi})^i,$ $Q$ is Carter constant. $\EuScript{N}_1$, $\EuScript{N}_2$, $\EuScript{N}_3$ are corresponding normalizing multipliers.
    
\begin{figure}
\includegraphics[width=0.7\linewidth]{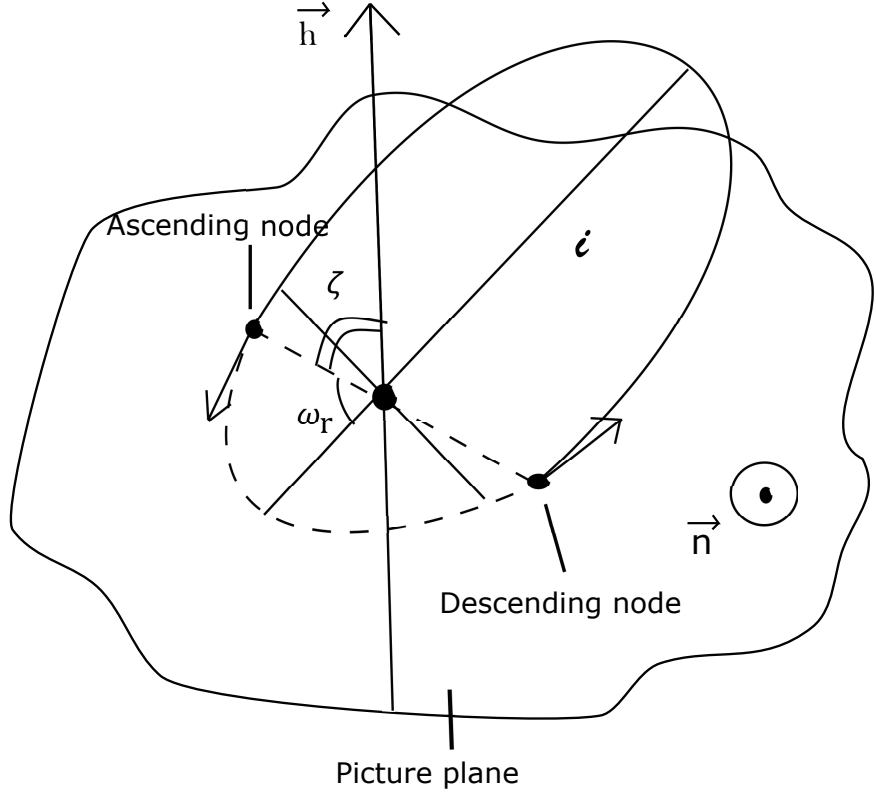}
\caption{Trajectory of the radius-vector of relative motion $x^{(\alpha)}$ in Fermi coordinates.}\label{Kepler1}
\end{figure}

In the linear order for $\rho/M,$ $v/c$ and neglecting of the time evolution of vector  $n^{(\alpha)}$ (relative to non rotating Fermi frame),  the expression $z_r(\tau)=\left(n_{(\alpha)}X_1^{(\alpha)}\right)'_{\tau}$ mathematically coincide with the expression of Doppler shift of light from a particle, slowly moving along a bound orbit near the mass $m=m_1 m_2/(m_1+m_2)$ on Kepler orbit. Consider the problem in this approximation. Fig. \ref{Kepler1} illustrates the meaning of the defined below parameters of the orbit of relative motion. Here we consider 3-dimensional space that is orthogonal to $h^i{}_{(4)}.$  

Denote the following abbreviations for the parameters of the orbit relative to the picture plane --- 2-dimensional plane that is orthogonal to vector $n^{(\alpha)}:$
\begin{itemize}
\item $i\in[0,\,\pi]$ --- orbital inclination is the angle between the picture plane and the plane of the orbit;
\item $\omega_r\in[0,\,\pi]$ --- pericenter longitude, counted along the orbital plane;
\item $\zeta\in[0.\,2\pi]$ --- position angle, counted along the picture plane.
\end{itemize}  
Period of the relative motion of the components can be found with very good accuracy by considering the function $(1+z(\tau))/f(t(\tau))$ that is approximately periodic:
\begin{equation}
T\approx \frac{\Delta T_N}{(N-1)}\,
\end{equation}
where $\Delta T_N$ is the interval of proper time that is bounded by local maximums of the function and consist of $N$ maximums.
Consider function $z_r(\tau)$ through time interval $T.$ Denote the local maximum of the function $z_r(\tau)$ as $z_a$ (corresponding proper time $\tau_a$) and the following local minimum as $z_c$ (corresponding proper time $\tau_c$). Then $\tau_a<\tau_c$. Also denote times $\tau_b,$ $\tau_d$ such that $z_r(\tau_b)=0,$ $z_r(\tau_d)=0,$  $\tau_a<\tau_b<\tau_c<\tau_d$ and $\tau_d-\tau_a=T.$ Define
\begin{equation}
h_1=\int\limits^{\tau_b}_{\tau_a}z_r(\tau)\mathrm{d}\tau\,;\quad h_2=-\int\limits^{\tau_d}_{\tau_c}z_r(\tau)\mathrm{d}\tau\,.
\end{equation}  
Using the Lemann-Files method (see, e. g. \cite{Cherepaschuc}), obtain:
\begin{align}
\left\{
\begin{aligned}
&\omega_r=\arctg1\left[\frac{2\sqrt{-z_a z_c}}{z_a-z_c}\frac{(h_1-h_2)}{(h_1+h_2)}\right]\,;\notag\\
&e=\frac{z_a+z_c}{z_a-z_c}\frac{1}{\cos\omega_r}\,.\label{Newtonpar}
\end{aligned}
\right.
\end{align}
Here $e$ is the eccentricity of the orbit. In the Newtonian limit it is possible to find only one more parameter of motion --- mass function $M_2$:
\begin{equation}\label{M2N}
M_2=\frac{m_2 \sin i}{(m_1+m_2)^{2/3}}=c\left(\frac{T}{16\pi G}\right)^{1/3}(z_a-z_b)\sqrt{1-e^2}\,.
\end{equation}

The presence of external gravitational field decrees the symmetry of the system relative to the considered approximation (Newtonian motion in the flat space-time). Due to this it is possible to anticipate that the addition parameters can be found as a result of more detailed investigation of the inverse problem for the relative motion.  

In the 3-dimensional hypersurface that is orthogonal to $h^i{}_{(4)}$ define the unit vector $a^{(\alpha)}$ that is orthogonal to the plane of the orbit of the relative motion and the vector $b^{(\alpha)}$ that is parallel to the main axis of the orbit and has direction from the center of mass to the orbit pericenter. Consider the relative motion on the time scale more than $T$ but comparable to $T_0.$ Orbit of relative motion rotates with angular velocity $\omega^{(\alpha)}.$ Then obtain the following relations
\begin{align}
\left\{
\begin{aligned}
&n^{(\alpha)}a_{(\alpha)}=\cos i\,;\\
&\dot{\cos i}=e^{(\alpha)(\beta)(\gamma)}n_{(\alpha)}\omega_{(\beta)}a_{(\gamma)}+\dot{n}^{(\gamma)}a_{(\gamma)}\,;\\
&b_{(\alpha)}a^{(\alpha)}=0\,;\label{Relative}\\
&\dot{(\sin i\sin\omega_r)}=-e^{(\alpha)(\beta)(\gamma)}n_{(\alpha)}\omega_{(\beta)}b_{(\gamma)}-\dot{n}^{(\gamma)}b_{(\gamma)}\,.
\end{aligned}
\right.
\end{align}
Here dot denotes derivative with respect to proper time. It can be approximately expressed from residuals. For example 
\begin{equation}
\dot{\cos i}\approx \frac{\cos i(\tau+T)-\cos i(\tau)}{T}\,.
\end{equation}
The values for $\omega_r$ are found with periodicity $T$ from Lemann-Files method and $\sin i$ can be expressed as $\sin i=M_2/W,$ where $W$ is a constant of motion:
\begin{equation}\label{defW}
W=m_2/(m_1+m_2)^{2/3}\,.
\end{equation}
From (\ref{Relative}) follows the following bisquare equation for $W:$
\begin{align}\label{mainW}
&-\cos^2\omega_r B_{(\alpha)}B^{(\alpha)}W^4+\notag\\
&\left[\cos^2\omega_r((\dot{M_2})^2+B_{(\alpha)}B^{(\alpha)}(M_2)^2)+(M_2)^2(\dot{\sin\omega_r})^2\right]W^2-(M_2)^4(\dot{\sin\omega_r})^2=0\,.
\end{align}
Here $B^{(\alpha)}$ defined by (\ref{B}). It must be calculated at arbitrary time $\tau$ that is belong to considered interval $[T_1+n_1T,\,T_1+n_1T+T].$
 The equation (\ref{mainW}) (with residuals instead of differentials) can be solved for $N-1$ pair of the values of proper time: $\left\{T_1;\,T_1+T\right\},\,...$ $\left\{T_1+(N-1)T;\,T_1+NT\right\}\,.$ For each such pair the equation (\ref{mainW}) has two different positive solutions $W$ in general case. This is unlikely that they are coincide for different intervals. Due to this it is possible to use this as criterion for the chose of the true root that is not changes with time $W=M_2/\sin i.$ In order to maximise the accuracy of calculation it is necessary to chose average value of all solutions $W$ as resulting parameter.
  
It is possible to find also orbital inclination:
\begin{equation}
\sin i=\frac{M_2}{W}\,.
\end{equation}
In the considered approximation the problem is symmetric under the direction of rotation. Due to this it is possible to obtain only $\sin i$ not $i.$ This is equivalent to the consideration only the interval $i\in[0,\pi/2].$

In order to determine the position angle $\zeta$ it is necessary to determine the direction $\vec h$ in picture plane for the measurement of the angle $\zeta$ from this direction (see Fig. \ref{Kepler1}). It is convenient to choose this direction as $h^{(\alpha)}=e^{(\alpha)(\beta)(\gamma)}n_{(\beta)}h_{1(\gamma)}.$ Then the 3 dimension vector $\vec h$ has the sense of orthogonal to both ray vector $n^{(\alpha)}$ and the azimuthal direction of rotating black hole. We obtain:
\begin{align}\label{posangle}
\sin\zeta=\frac{\dot{(\cos i)}e^{(\alpha)(\beta)(\gamma)}h_{1(\alpha)}B_{(\gamma)}n_{(\beta)}+B^{(\alpha)}h_{1(\alpha)}\sqrt{(\sin^2 i) B_{(\kappa)}B^{(\kappa)}-\dot{(\cos i)}^2}}{B_{(\kappa)}B^{(\kappa)}\sqrt{h_{1(\alpha)}h^{(\alpha)}{}_1-(h_{1(\alpha)}n^{(\alpha)})^2}}\,.    
\end{align}
Due to the symmetry relative to the direction of $\vec h$ it is possible to determine only the sinus of position angle (\ref{posangle}). For the unique solution we assume $\zeta\in[-\pi,\pi].$

As a result the influence of external gravitational field in the described approximation gives possibility to determine two more parameters of relative motion with respect to absence of the external gravitational field. Furthermore astrophysical model of the considered source of radiation can give possibility to determine its mass $m_1.$ The other case when the redshift of both stars components can be measured, then it is possible to determine the relation $m_1/m_2.$ In both cases the masses of the stars can be found using in addition the definition of $W$ (\ref{defW}).  
The results of calculating of parameters for an example of the described model are presented in Table \ref{relParameters}.
\begin{table}
\caption{The model parameters and the reconstructed from the solution of inverse problem parameters. The reconstruction is performed for the motion of the center of mass of the system for the interval of time $60<\tau<73$. The whole system of initial parameters of the system is presented also in Fig. \ref{tKerrRedShift_a01}}\label{relParameters}
\begin{tabular}{l c c}
 Parameter & Model value & Reconstructed value\\
\hline
  Eccentricity, $e$ & 0.68 & 0.58\\
  Period of relative motion, $T$ & 20.268 $Mc^{-1}$& 20.264 $Mc^{-1}$\\
  Mass function, $M_2$      & 0.0062 $M^{1/3}$ & 0.0066 $M^{1/3}$\\
   Pericenter longitude, $\omega_r$ & 1.56 rad& 1.32 rad\\
    Orbital inclination, $i$ & 1.55 rad& 1.31 rad\\
     Position angle, $\zeta$ & 1.72 rad& 3.05 rad\\
\hline
\end{tabular}
\end{table}

%

\begin{acknowledgements}
The authors would like to thank Alexander Tarasenko for useful discussions.
\end{acknowledgements}

\section{Conclusions}
The presented method gives possibilities for studying the motion of the binary star in strong external gravitational field of a rotating black hole using the value of redshift of spectral lines of received radiation. The consideration of the problem do not have restrictions on the motion of the binary as a whole and due to this it is possible to apply the method to the binaries that moves very close to supermassive black hole and therefore have a large velocity of the center of mass relative to distant observer. In this paper the possibility of application of the method is shown for the case of external gravitational field of Kerr black hole. But the method without difficulties can be generalised for the cases of other strong external gravitational fields.    

For the solution of direct problem (calculation of redshift of light that is emitted by the star in binary) the more accurate formulas have been used than in the case of inverse problem. The main approximation is the consideration of the picture plane of relative motion of the stars as covariantly constant on time internal of $\sim T$ (this is true in good approximation if the black hole is sufficiently massive). Due to this the reconstruction of parameters of relative motion is performed with certain errors (see Table \ref{relParameters}).  The largest error in the parameter $\zeta$ is due to the approximate axial symmetry of the problem relative to rotations in picture plane. The possibility of reconstruction of the trajectory of motion of components open a prospect for investigation of such characteristic of relative motion as precession relative to the inertial reference frame due to the influence of external gravitational field. Or precession and deformation of the orbit of relative motion of the stars due to the gravitational interactions with external gravitational field. All these effects can be estimate using the reconstructed parameters of the orbit for different intervals of time of observation.   

Due to the simple relation between the redshift and the duration of time of arrival of pulses of pulsar (see, e. g., \cite{Zhang2017}), the method described in the present paper can be applied not only to studying the redshift of registered spectral lines of the star, but also for analysing pulsar timing data in the case of the motion of a binary pulsar in a strong external gravitational field.   




\end{document}